\documentclass[10pt, aps, prb,twocolumn,superscriptaddress]{revtex4-1}

\usepackage{graphicx}
\usepackage{hyperref}
\usepackage[all]{hypcap}
\usepackage{amsmath}
\usepackage{bm}
\usepackage{etoolbox}
\hypersetup{colorlinks=true,linkcolor=blue,citecolor=blue,urlcolor=blue}

\newcommand{\mytilde}{\raise.17ex\hbox{$\scriptstyle\mathtt{\sim}$}}
\newcommand{\UBC}{Stewart Blusson Quantum Matter Institute, University of British Columbia, Vancouver, Canada V6T 1Z1}
\newcommand{\USASK}{Department of Physics \& Engineering Physics, University of Saskatchewan, Saskatoon, Canada S7N 5E2}
\newcommand{\STUT}{Max Planck Institute for Solid State Research, Heisenbergstra\ss e 1, 70569 Stuttgart, Germany}
\newcommand{\WURZ}{Experimentelle Physik IV and R\"ontgen Research Center for Complex Materials (RCCM), Fakult\"at f\"ur Physik und Astronomie, Universit\"at W\"urzburg, Am Hubland, D-97074 W\"urzburg, Germany}
\newcommand{\TWEN}{Faculty of Science and Technology and MESA+ Institute for Nanotechnology, University of Twente, 7500 AE Enschede, The Netherlands}
\newcommand{\CLS}{Canadian Light Source, University of Saskatchewan, Saskatoon, SK S7N 2V3, Canada}

\newcommand{\DRES}{Institute for Structure Physics, Dresden Technical University, 01062 Dresden, Germany}
\newcommand{\CHN}{National Synchrotron Radiation Laboratory, University of Science and Technology of China, Hefei 230026, Anhui, China}

\begin{document}

\title{Intrinsic versus extrinsic orbital and electronic reconstructions at complex oxide interfaces}

\author{R. J. Green}
\email[]{robert.green@usask.ca}
\affiliation{\UBC}
\affiliation{\USASK}

\author{V. Zabolotnyy}
\affiliation{\WURZ}

\author{M. Zwiebler}
\affiliation{\DRES}

\author{Z. Liao}
\affiliation{\TWEN}
\affiliation{\CHN}

\author{S. Macke}
\affiliation{\UBC}
\affiliation{\STUT}

\author{R. Sutarto}
\affiliation{\CLS}

\author{F. He}
\affiliation{\CLS}

\author{M. Huijben}
\affiliation{\TWEN}

\author{G. Rijnders}
\affiliation{\TWEN}

\author{G. Koster}
\affiliation{\TWEN}

\author{J. Geck}
\affiliation{\DRES}

\author{V. Hinkov}
\affiliation{\WURZ}

\author{G. A. Sawatzky}
\affiliation{\UBC}

\date{\today}

\begin{abstract}
The interface between the insulators LaAlO$_3$ and SrTiO$_3$ accommodates a two-dimensional electron liquid (2DEL)---a high mobility electron system exhibiting superconductivity as well as indications of magnetism and correlations. While this flagship oxide heterostructure shows promise for electronics applications, the origin and microscopic properties of the 2DEL remain unclear. The uncertainty remains in part because the electronic structures of such nanoscale buried interfaces are difficult to probe, and is compounded by the variable presence of oxygen vacancies and coexistence of both localized and delocalized charges. These various complications have precluded decisive tests of intrinsic electronic and orbital reconstruction at this interface. Here we overcome prior difficulties by developing an interface analysis based on the inherently interface-sensitive resonant x-ray reflectometry. We discover a high charge density of 0.5 electrons per interfacial unit cell for samples above the critical LaAlO$_3$ thickness, and extract the depth dependence of both the orbital and electronic reconstructions near the buried interface. We find that the majority of the reconstruction phenomena are confined to within 2 unit cells of the interface, and we quantify how oxygen vacancies significantly affect the electronic system. Our results provide strong support for the existence of polarity induced electronic reconstruction, clearly separating its effects from those of oxygen vacancies.
\end{abstract}

\maketitle

The complex interplay between charge, spin, orbital, and lattice degrees of freedom in transition metal oxides leads to a multitude of emergent phenomena having numerous potential applications.\cite{Khomskii2014book} Often the phenomena are realized through explicit fine tuning of these degrees of freedom, for example via electronic doping or applied pressure. Recent advances in atomic scale oxide thin film growth have yielded a new method of precisely tuning material properties via heterostructuring, where sequences of films are grown epitaxially on top of specially chosen substrates.\cite{Mannhart_Science_2010, Hwang_NatMat_2012} Orbital, spin, and electronic reconstructions which can occur at the atomically precise interfaces lead to emergent properties that are often very different from the corresponding bulk materials.\cite{Hwang_NatMat_2012}

The interface between LaAlO$_3$ and SrTiO$_3$ (LAO/STO) is paradigmatic of the oxide heterostructure field, hosting a two dimensional electron liquid (2DEL) \cite{Hwang_Nature_2004,Thiel_Science_2006} that exhibits gate-tunable superconductivity,\cite{CavigliaPRL2019,LevyPRL2018,FernandesPRL2018,CavigliaNatComm2018,Reyren_Science_Super_2007,Li_NatPhys_MagSup_2011} flexoelectricity,\cite{DaiPRL2019} magnetism,\cite{Pryds2020NatComm,Brinkman_NatMat_2007,Bert_NatPhys_MagSup_2011} and correlations.\cite{Breitschaft_2DEL_PRB_2010,KhannaPRL2019} Recent experiments have further utilized this interface as a foundation for the construction of one-dimensional superconducting electron waveguides \cite{LevyPRL2018,Levy2020SciAdv,Levy2020Science} which have potential application in quantum computing.  However, while the system exhibits such fascinating properties and shows such promise for device applications,\cite{Mannhart_Science_2010} uncertainty remains regarding the origin and physical details of this highly studied interfacial electronic liquid.\cite{Salluzzo2020PRL} Some of the earliest studies of the interface 2DEL \cite{Nakagawa2006NatMat} proposed that it results from an electronic reconstruction due to a \emph{polar catastrophe}---that a diverging electric potential originating from charged LaO$^+$ and AlO$_2^-$ layers is rectified through a spontaneous relocation of 0.5 e$^-$ per square unit cell (u.c.) from the LAO surface to the interface for heterostructures with LAO thicknesses greater than 3 u.c., thus forming the 2DEL. The concept of electronic reconstruction had been introduced several years earlier in studies of polar surfaces.\cite{Hesper2000} Crucially, however, transport measurements on LAO/STO typically find about an order of magnitude less charge than expected from electronic reconstruction.\cite{Thiel_Science_2006} More recent studies have shed some light on this issue, showing that a certain amount of charge is localized and thus hidden to transport measurements.\cite{Sing_PRL_HXPS_2009,2014_Rusydi_Natcom}

\begin{figure*}
\includegraphics[width=140mm]{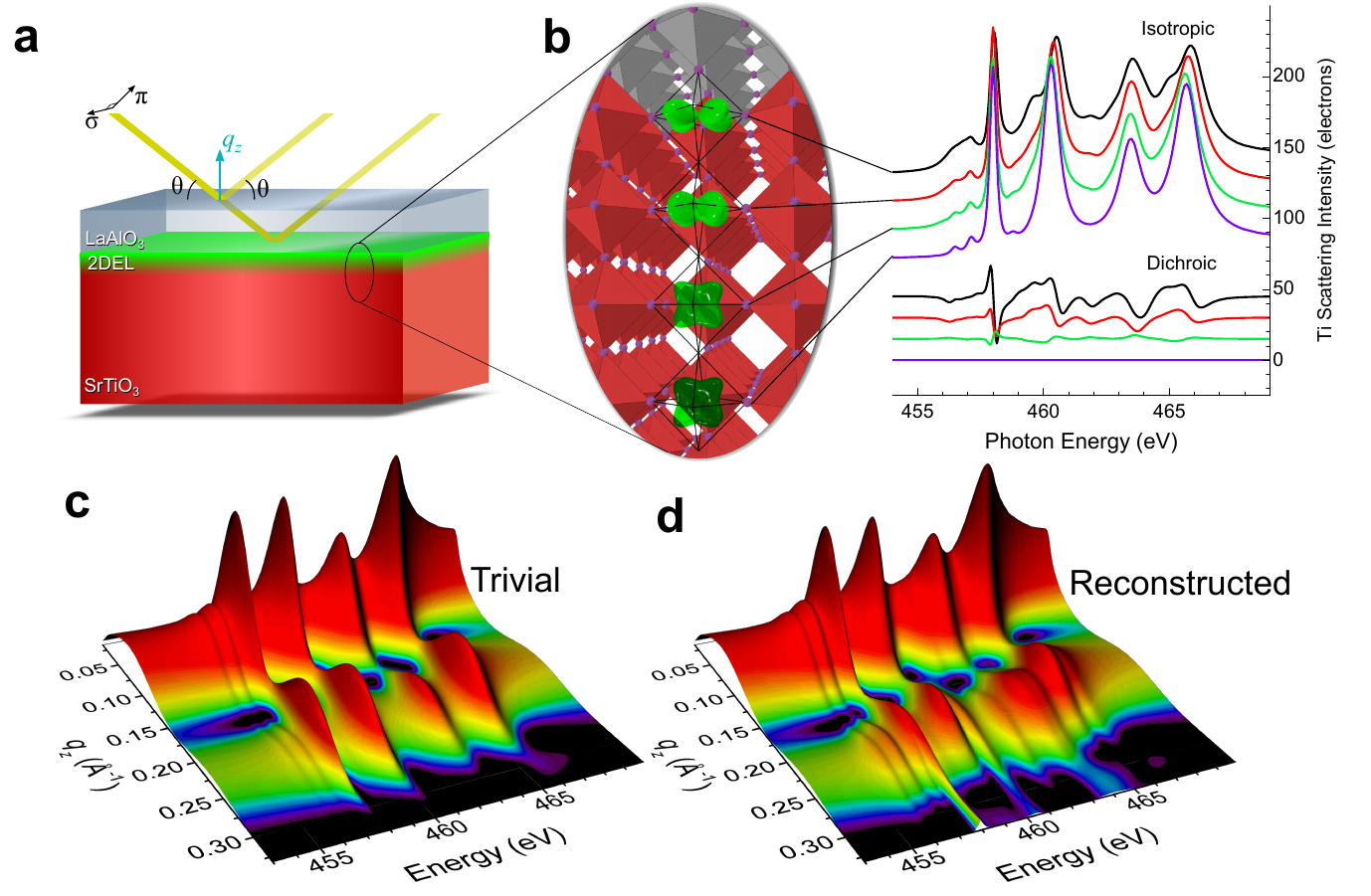}
\caption{\textbf{Theoretical resonant reflectometry of LaAlO$_3$/SrTiO$_3$ heterostructures.} \textbf{a}, Experimental geometry for the reflectometry measurements, where the photon polarization is either within ($\pi$) or perpendicular to ($\sigma$) the scattering plane. Specific geometries are characterized by either the reflection angle $\theta$ or the photon momentum transfer in the $z$ direction, $q_z$. \textbf{b}, The graphic depicts orbital and electronic reconstructions at the interface via the orbital symmetry and opacity, respectively. For each atomic layer with a given orbital occupation, there exist isotropic and dichroic resonant scattering responses shown, which then combine to determine the reflectivity response of the material. \textbf{c-d}, Simulated reflectivity maps for $\pi$-polarized light and energies spanning the Ti $L_{2,3}$ resonance on a 10 u.c. LAO/STO heterostructure in the case of no reconstructions (\textbf{c}), and combined orbital and electronic reconstructions (\textbf{d}).}
\label{Fig:1}
\end{figure*}

In addition to the total charge and its confinement profile near the interface, a crucial characteristic of the 2DEL is its orbital symmetry, which directly controls the mobility of the electrons and contributes to the energetic landscape of magnetism and superconductivity.\cite{FernandesPRL2018,CavigliaPRL2019} The interface electrons reside in the Ti $3d$ $t_{2g}$ band, which is split into $d_{xy}$ and $d_{xz/yz}$ subbands due to the tetragonal symmetry breaking imposed by the interface. First principles calculations predict that the $d_{xy}$ bands are lower in energy near the interface and thus host the majority of the charge, whereas $d_{xz/yz}$ bands become more occupied further from the interface \cite{Popovic_PRL_origin_2008, SonPRB2009, StemmerReview2014}. X-ray absorption spectroscopy (XAS) experiments have verified this qualitatively, as linear dichroism measurements agree with the presence of lower energy $d_{xy}$ orbitals.\cite{Salluzzo_PRL_2009} However, the magnitude of this orbital reconstruction, and its spatial dependence moving away from the interface, have yet to be determined.

Here we study the electronic structure of the LAO/STO interface using synchrotron-based resonant x-ray reflectometry (RXR). RXR is the union of two powerful, complementary techniques: x-ray absorption spectroscopy, which probes the element-specific electronic structure (in particular the local $3d$ orbital occupations) through core-valence resonant electronic excitation, and reflectivity, which inherently contributes excellent interface and depth sensitivity while also providing a natural quantitative detection scheme.\cite{Macke2014AdvMat, MackeGoering_2014} This combination of capabilities allows us to study the microscopic electronic structure properties specific to the 2DEL at the interface in exceptional detail. Previous studies on LAO/STO superlattices utilized a qualitative analysis of limited datasets to study electronic and orbital reconstructions.\cite{WadatiRXR2009, ParkRXR2013} Here we leverage the full power of RXR via a fully quantitative analysis of large data sets. In an extensive set of experiments, we apply RXR to LAO/STO samples below and above critical thickness for 2DEL formation, and before and after annealing in an $\mathrm{O_2}$ environment to eliminate potential oxygen vacancies, showing how orbital and electronic reconstruction define the spontaneous 2DEL formation.

\begin{figure*}
\includegraphics[width=183mm]{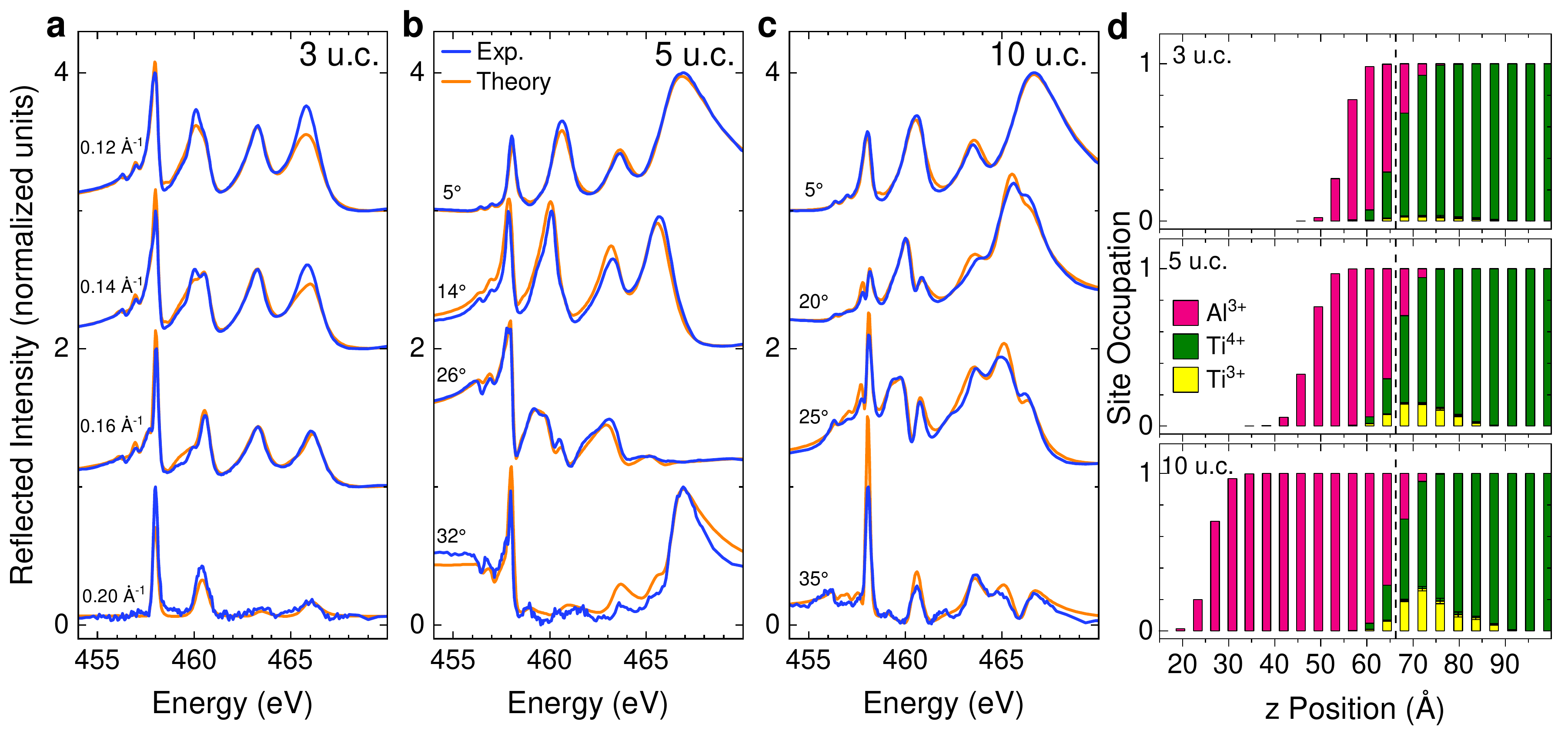}
\caption{\textbf{Resonant reflectometry data and extracted charge profile for as-grown LaAlO$_3$/SrTiO$_3$.} \textbf{a-c}, Experimental and modelled $\pi$-polarized energy scans across the Ti resonance at indicated fixed momentum transfer values or reflection angles for 3 u.c. (\textbf{a}), 5 u.c. (\textbf{b}), and 10 u.c. (\textbf{c}) samples. \textbf{d}, B-site cation concentrations and 2DEL charge depth profile near the interface.  Each vertical bar represents an atomic layer, so the distance between each pair is one unit cell. The dashed vertical lines denote the approximate interface where equal amounts of Ti and Al are found. Extended data sets and simulation results can be found in the supplementary material.}
\label{Fig:2}
\end{figure*}

In Figure \ref{Fig:1}(a) we depict the RXR experiment, showing the typical geometry for our reflection measurements using linearly polarized synchrotron x-rays. By changing the reflection angle or energy, one changes the momentum transfer $q_z$, leading to different interference effects (partial cancellation or enhancement of waves reflecting at different depths) and different interface sensitivity. Figure \ref{Fig:1}(b) shows via simulations the origin of reconstruction sensitivity in RXR. For the atomic planes near the interface exhibiting a particular reconstruction effect (where in the inset we depict orbital reconstructions via schematic orbital symmetries and electronic reconstructions via the opacity of the orbitals), the Ti atoms have very specific and different resonant scattering characteristics. Both the isotropic and linear-dichroic responses of the scattering show strong dependences on orbital occupation at these Ti $L_{2,3}$ resonance energies.

The resonant atomic scattering responses combine to yield the electric susceptibility of the material, and thus a particular reconstruction at the interface leads to a specific reflectivity response, which we investigate with model data in Fig. \ref{Fig:1}(c-d). Here the predicted RXR for a 10 u.c. LAO/STO sample is shown for energies spanning the Ti $L_{2,3}$ resonance and reflection angles spanning a typical measurable range.\footnote{Note we multiply the reflectivity spectra here by $q_z^4$ in order to show a wide $q_z$ range where the intensity spans several orders of magnitude}  As shown by the two plots, at high momentum transfers, very pronounced effects are expected in the RXR data due to orbital and electronic reconstructions, even if they are confined to only 2 u.c. near the buried interface as in the realistic example simulated here. This strong sensitivity, and the quantitative nature of RXR, makes this technique very useful for studying oxide interfaces.

We have applied RXR to a series of samples to perform a comprehensive study of the LAO/STO reconstruction phenomena. Figure \ref{Fig:2} details the reflectivity data and analysis results for samples having LAO thicknesses of 3, 5, and 10 u.c. Fig. \ref{Fig:2}(a-c) show selected reflectivity data along with model fit results (complete datasets are presented in the supplemental material). Signatures of the thickness-dependent onset of the 2DEL are immediately evident from the data, as the 3 u.c. sample exhibits a less featured reflectivity response similar to Fig. \ref{Fig:1}(c), whereas the spectra of the thicker samples reveal a much more detailed structure, comparable to Fig. \ref{Fig:1}(d). 

\begin{figure}
\includegraphics[width=89mm]{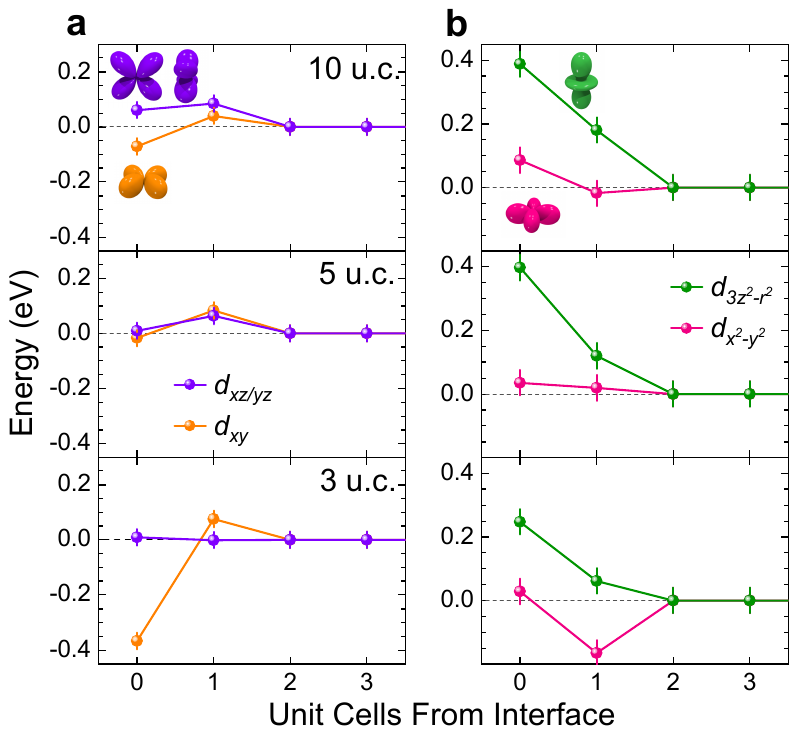}
\caption{\textbf{Orbital reconstruction of as-grown LAO/STO quantified via reflectometry.} \textbf{a}, $t_{2g}$ and \textbf{b}, $e_g$ orbital energies for the Ti atomic layers near the LAO/STO interface, relative to the bulk orbital energies. }
\label{Fig:3}
\end{figure}

We model the depth-dependent electric susceptibility of the heterostructures starting from the atomic level, where the Ti resonant scattering response is computed using quantum many-body theory \cite{Haverkort_Wannier_PRB_2010, QuantyWeb} while the off-resonant scattering response of all atoms are taken from tabulated data.\cite{Henke} The resonant response depends strongly on the local $3d$ orbital occupation of the Ti atoms, thus providing the sensitivity to the 2DEL density and symmetry. From the susceptibility we compute the energy- and angle-dependent reflectivity response, and fit the material parameters as described in the Methods section. The fit results are overlaid with the experimental data in Fig. \ref{Fig:2}(a-c), where an excellent level of agreement is evident. The models corresponding to these fits, which quantify the interfacial reconstruction phenomena, are shown in Fig. \ref{Fig:2}(d) and Fig. \ref{Fig:3}. In Fig. \ref{Fig:2}(d) we plot fractional atomic site occupations for the cations in the samples (full concentration profiles, including anions, are presented in the supplemental material), where our resonant Ti analysis allows us to distinguish between Ti$^{4+}$ and Ti$^{3+}$ valences (i.e. the respective absence or presence of the 2DEL charge). Evident is the $\sim$1-2 u.c. average roughness at both the surface and the interface, comparable to which has been found with other techniques.\cite{YacobyPRL2007, Nakagawa2006NatMat, Chambers2010} Strong evidence is present for the the 4 u.c. threshold for 2DEL formation, as the 3 u.c. sample is found to have significantly less charge than the thicker samples. As shown in Table \ref{Tab:Charge}, for the 3 u.c. sample our fit converges with a small amount of charge at 0.16 e$^-$/u.c. This charge is distributed rather evenly over 4 nm near the interface, whereas the thicker samples have much more pronounced concentrations confined to the interface. The total charge quantities in the 5 and 10 u.c. samples are significant, as also shown in Table \ref{Tab:Charge}. 

\begin{table}
{
\begin{ruledtabular}
\begin{tabular}{ cc } 
~~~~Sample~~~~ & ~~~~Charge (e$^-$/u.c.)~~~~ \\ \hline
~3 u.c.  & 0.16 $\pm$ 0.03 \\
~5 u.c.  & 0.59 $\pm$ 0.05 \\
10 u.c.  & 0.97 $\pm$ 0.08 \\
~~~~~~~~~~~~~~~10 u.c. (annealed)  & 0.50 $\pm$ 0.03 \\
\end{tabular}
\end{ruledtabular}
\caption{\label{Tab:Charge} Total charge hosted in the Ti $3d$ orbitals near the interface for each sample, as extracted from the RXR fits. }
}
\end{table}

In Fig. \ref{Fig:3}, the extracted interface induced symmetry breaking of the Ti 3d orbitals is detailed. While the 3d orbitals have $O_h$ symmetry in the bulk of the STO, the presence of the interface leads to a tetragonal local symmetry distortion.\cite{Salluzzo_PRL_2009}  By computing the tetragonally symmetric Ti resonant scattering tensor according to the energies of the 3d orbitals, and fitting to our polarization-dependent RXR spectra, we can extract the depth dependence of the orbital energies relative to those of the bulk.  For all samples (regardless of 2DEL presence), we detect a strong symmetry breaking in the 2 u.c. closest to the interface.  We find the $d_{xy}$ orbital to be lowest in energy near the interface, as predicted by first principles modelling \cite{Popovic_PRL_origin_2008, SonPRB2009} and qualitatively observed experimentally.\cite{Salluzzo_PRL_2009} Further, we detect that the tetragonal distorting of orbital energies is enhanced in the presence of the 2DEL, a feature also previously predicted from first principles \cite{StemmerReview2014}. Given the different effective masses for $d_{xy}$ versus $d_{xz/yz}$ carriers, a detailed quantification of this orbital symmetry breaking is key to understanding the comprehensive transport properties of the LAO/STO system.

\begin{figure}
\includegraphics[width=89mm]{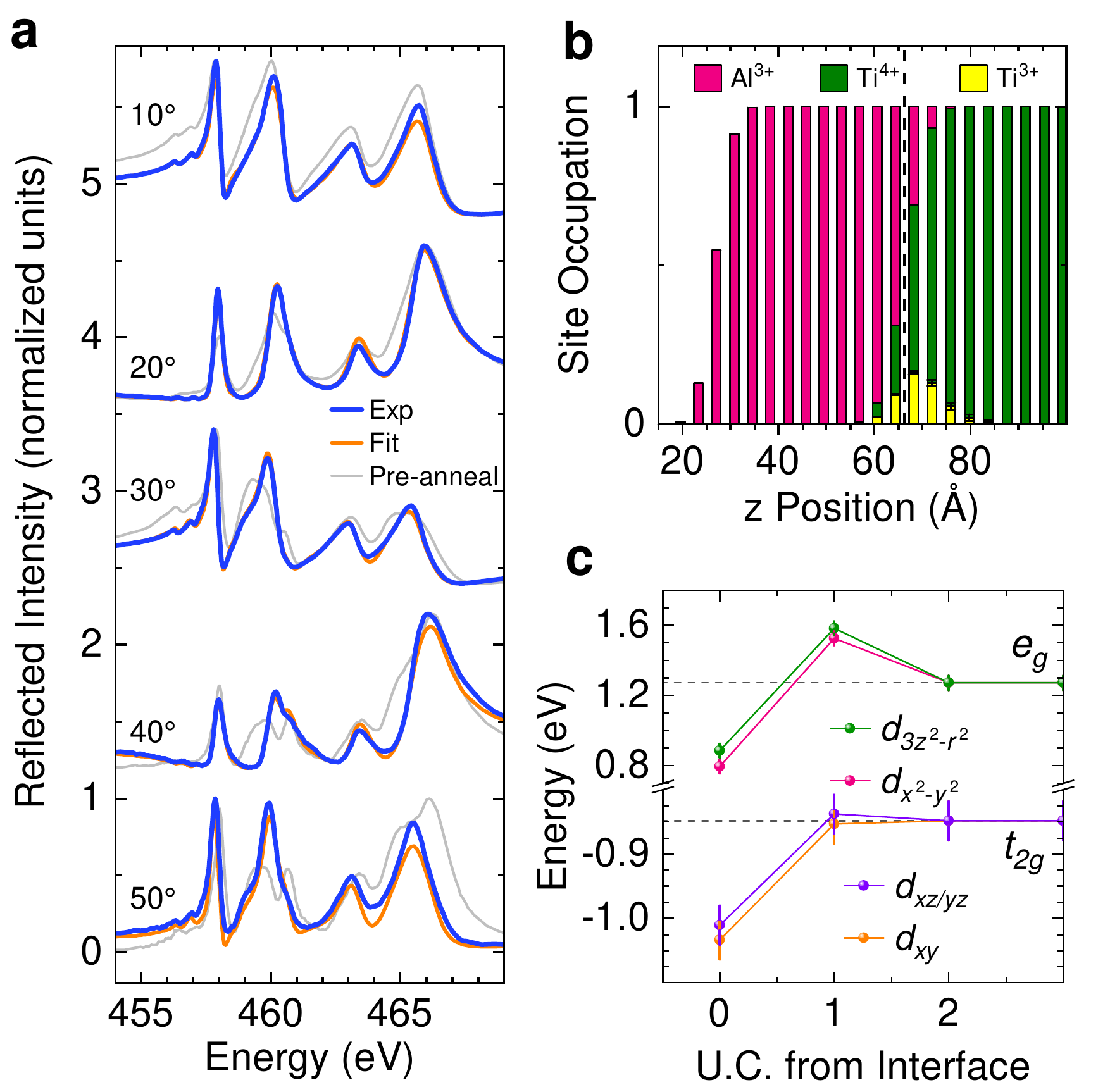}
\caption{\textbf{Effects of post-growth oxygen annealing.} \textbf{a}, Experimental and modelled fixed angle, $\sigma$-polarized, reflectometry scans across the Ti resonance are shown for a 10 u.c. LAO/STO after annealing in an O$_2$ environment. The experimental data for the same sample before annealing are shown in grey for comparison. \textbf{b}, 2DEL charge profile after annealing. \textbf{c}, Orbital energies after annealing, extracted from the reflectometry analysis. }
\label{Fig:4}
\end{figure}

While a common observation in many experimental studies is that the 2DEL total charge is much less than the 0.5 e$^-$ per square u.c. predicted by the polar catastrophe model,\cite{Thiel_Science_2006} we find here quantities relatively close to this value---in fact they are systematically slightly higher. Oxygen vacancies (either in the LAO film or the STO substrate) can be an electron dopant, leading to excess carriers in the 2DEL as well as electron trapping.\cite{AartsPRL2020} To investigate the possible effect of oxygen vacancies, we measured an RXR dataset on the same 10 u.c. sample, immediately after annealing in a partial pressure (200 mbar) O$_2$ environment. The results, shown in Fig. \ref{Fig:4}, display a significant change in the RXR response upon annealing, indicating a strong sensitivity of RXR to the oxygen content and corresponding Ti charge variation. The data and model fits are shown in Fig. \ref{Fig:4}(a), with the pre-annealing data plotted as well for comparison. Performing the same quantitative analysis for this case, we find that the charge density is reduced from \mbox{0.97 e$^-$/u.c.} down to a post-annealed value of \mbox{0.50 $\pm$ 0.03 e$^-$/u.c.,} in agreement with the predicted value of \mbox{0.5 e$^-$/u.c.} within our fitting uncertainty. This decrease in 2DEL total charge is consistent with what we observe in transport measurements (see supplemental material). The charge profile is plotted in Fig. \ref{Fig:4}(b), where it is evident the total charge decreases post-annealing and also becomes more confined to the interface (without considering interface broadening, the mean 2DEL charge depth is 0.4 u.c. past the first TiO$_2$ layer post-annealing versus 1.5 u.c. before).  Finally, in Fig. \ref{Fig:4}(c) we show the post-annealing orbital energy profiles. Interestingly, the orbital energies after annealing exhibit a much weaker tetragonal splitting (23 versus 133 meV), suggesting that the larger orbital splittings for the 10 u.c. sample prior to annealing in Fig. \ref{Fig:3} could be due primarily to strong potential variations due to oxygen vacancies.\cite{PavlenkoPRB2012} Further, the $d_{xy}$ orbital is shifted to a lower energy at the interface post-annealing, explaining the tighter confinement of the 2DEL charge.

Our resonant x-ray reflectivity analysis of LaAlO$_3$/SrTiO$_3$ 2D electron liquids provides crucial answers to questions regarding oxide interface reconstruction phenomena. We find total charge quantities consistent with the theory of polarity-induced electronic reconstruction, elucidate the effects of oxygen vacancies on this system, and further extract the confinement profile of the reconstructed charge. Additionally, the RXR technique allows us to quantify the orbital symmetry of electronic states at Angstrom length scales, showing that significant symmetry breaking effects of the Ti $3d$ orbitals are confined to within 2 unit cells of the interface. Crucially, the nature of RXR allows us to probe the entire reconstructed charge density, not just mobile carriers. Building upon the results presented here, future RXR experiments will be helpful to closely examine the 4 u.c. threshold for 2DEL formation, as well as O$_2$ annealing of samples of various thicknesses. Additionally, with such a precise quantification of the 2DEL characteristics, future efforts can be directed toward engineering these properties for the development of functional devices, either in the LAO/STO system, or other oxide heterostructures. Finally, while these findings are important for the understanding of the LAO/STO electronic system and related oxide reconstruction phenomena, our results also show resonant x-ray reflectivity to be an invaluable tool for the burgeoning field of oxide interfaces.

\noindent\textbf{Methods}

\begin{footnotesize}
The set of samples investigated included 3, 5, and 10 u.c. LaAlO$_3$ films grown by pulsed laser deposition on TiO$_2$-terminated SrTiO$_3$ substrates.  A KrF excimer laser operating at 248 nm was used, with a fluence of 1 J/cm$^2$ and repetition rate of 1 Hz. The growth took place at 800 $^\circ$C in an O$_2$ pressure of $1\times 10^{-3}$ mbar. After deposition, all samples were cooled down in a higher O$_2$ pressure of $5\times 10^{-3}$ mbar at a rate of 30 $^\circ$C/min. Atomic layer growth was controlled using reflection high energy electron diffraction (RHEED), as reported in the supplemental material. Post annealed samples were annealed at a temperature of 530 $^\circ$C under an O$_2$ pressure of 200 mbar for 1 hour.

The resonant x-ray reflectometry experiments were carried out at the REIXS beamline of the Canadian Light Source,\cite{Hawthorn_REIXS} at 300K in an ultra-high vacuum environment ($<  10^{-9}$ mbar). Reflection intensity was monitored using a filtered photodiode, whose response function was measured using the direct synchrotron beam. Data was normalized by the incident beam flux and the photodiode response to obtain the quantitative reflectivity spectra. Two different measurement modes were used for data presented in this work.  For data in Fig. \ref{Fig:2}(a), the momentum transfer was held fixed by slightly varying the reflection angle for each energy.  For all other data the reflection angle was held fixed while varying incident energy. Either mode can be used for a successful RXR analysis, provided a sufficient sampling of the energy versus angle/momentum space is made.

Simulations of the Ti $L_{2,3}$ resonant response for given orbital energies and occupations were carried out with multiplet crystal field theory (MCFT). This model Hamiltonian approach includes the full multiplet effects arising from strong atomic Coulomb interactions, as well as crystal field perturbations arising from the bonding with oxygen. Exact diagonalization and simulations of the spectral functions were carried out with the software \emph{Quanty}.\cite{QuantyWeb, Haverkort_Wannier_PRB_2010, Haverkort_DMFT_2014, Haverkort_DMFTXAS_EPL2014} The simulated resonances were merged with tabulated off-resonant atomic scattering factors $f_2$,\cite{Henke} and the scattering factors $f_1$ were then calculated from the Kramers-Kronig relations. For all other elements (La, Sr, O, Al), off-resonant tabulated scattering factors were used. With all atomic scattering factors determined, the reflectivity was simulated and fit to experiment using the dynamical diffraction software QUAD.\cite{QUAD_PRL}
\end{footnotesize}

\medskip

\noindent\textbf{Acknowledgments}

\begin{footnotesize}
This work was supported by the Natural Sciences and Engineering Research Council of Canada, the Canadian Institute for Advanced Research, the Max Planck-UBC-UTokyo Centre for Quantum Materials, and the Canada First Research Excellence Fund, Quantum Materials and Future Technologies Program. Team members from the University of W\"urzburg were supported by the Deutsche Forschungsgemeinschaft (DFG), Project-ID 258499086-SFB 1170. Team members from the University of Twente were supported by the Netherlands Organization for Scientific Research (NWO). Part of the research described in this paper was performed at the Canadian Light Source, a national research facility of the University of Saskatchewan, which is supported by the Canada Foundation for Innovation (CFI), the Natural Sciences and Engineering Research Council of Canada, the National Research Council (NRC), the Canadian Institutes of Health Research (CIHR), the Government of Saskatchewan, and the University of Saskatchewan. We thank Research Computing at the University of Saskatchewan for computational resources.
\end{footnotesize}

\medskip

\noindent\textbf{Author contributions}

\begin{footnotesize}
Resonant reflectivity experiments were performed by R.J.G., S.M., R.S., and F.H. Data analysis was performed by R.J.G. with contributions from V.Z., M.Z., S.M., J.G., V.H., and G.A.S. Sample synthesis and initial characterization was managed and carried out by Z.L., M.H., G.R., and G.K. The manuscript was written by R.J.G. with input from all authors.
\end{footnotesize}

%temp packages
\newcommand\FramedBox[3]{%
  \setlength\fboxsep{0pt}
  \fbox{\parbox[t][#1][c]{#2}{\centering\huge #3}}}

\renewcommand*{\citenumfont}[1]{S#1}
\renewcommand*{\bibnumfmt}[1]{[S#1]}

\makeatletter 
\renewcommand{\thefigure}{S\@arabic\c@figure}
\makeatother

\makeatletter 
\renewcommand{\thetable}{S\@arabic\c@table}
\makeatother

\makeatletter 
\renewcommand{\theequation}{S\@arabic\c@equation}
\makeatother

%\makeatletter
%\apptocmd{\thebibliography}{\global\c@NAT@ctr 39\relax}{}{}
%\makeatother

\setcounter{equation}{0}
\setcounter{figure}{0}
\setcounter{table}{0}
\setcounter{page}{0}

\widetext
\clearpage

\begin{center}
\textbf{\emph{Supplementary Information for} ``Intrinsic versus extrinsic orbital and electronic reconstructions at complex oxide interfaces''}
\end{center}

\section{Sample Synthesis and Characterization}

\begin{figure*}[b]
\begin{center}
\includegraphics[width=140mm]{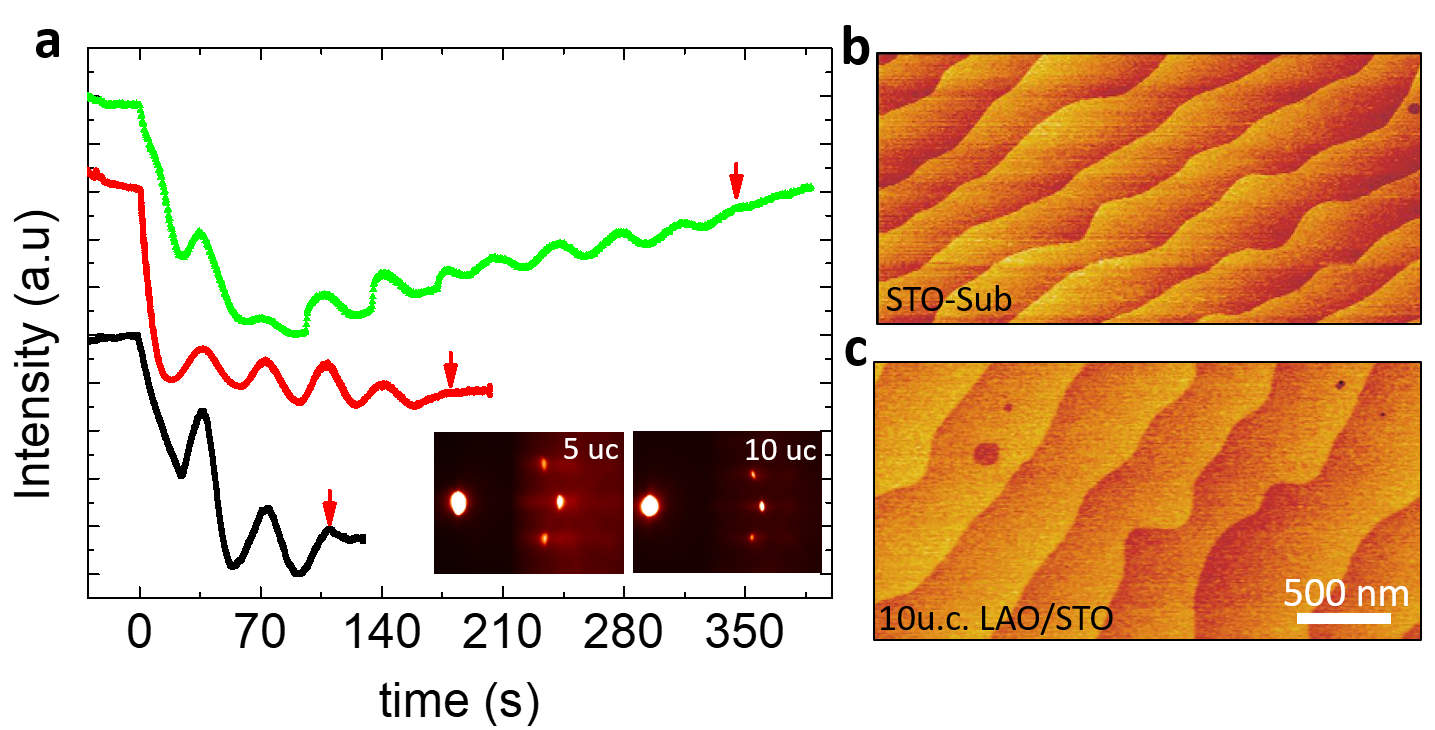}
\caption{\textbf{Film growth and characterization.} \textbf{a}, RHEED oscillations verifying a layer by layer growth for the 3, 5, and 10 u.c. LAO films grown on STO. The red arrows indicate when the growth stops. \textbf{b-c}, AFM image of (\textbf{b}) an STO substrate and (\textbf{c}) a 10 u.c. LAO/STO heterostructure showing large atomic-layer terraces.}
\label{Fig:Growth}
\end{center}
\end{figure*}

The LaAlO$_3$ films studied in this work were grown on TiO$_2$-terminated SrTiO$_3$ substrates using pulsed laser deposition. A KrF excimer laser was used, operating at 248 nm with a fluence of 1 J/cm$^2$ and repetition rate of 1 Hz. The substrate temperature during growth was 800$^\circ$C and an O$_2$ atmosphere of $1\times 10^{-3}$ mbar was maintained. After growth, the films were cooled to room temperature at a rate of 30 $^\circ$C/min under a higher O$_2$ pressure of $5\times 10^{-3}$ mbar.   The layer-by-layer growth of the films was confirmed by RHEED measurements, as shown in Fig. \ref{Fig:Growth}\textcolor{blue}{a} for the 3, 5, and 10 u.c. films.  The high quality of the SrTiO$_3$ substrates is demonstrated by the AFM image in Fig. \ref{Fig:Growth}\textcolor{blue}{b}, where atomic terraces are evident. This high quality is retained after the layer by layer LAO growth, as indicated in Fig. \ref{Fig:Growth}\textcolor{blue}{c} for a 10 u.c. sample.

\begin{figure*}
\begin{center}
\includegraphics[width=140mm]{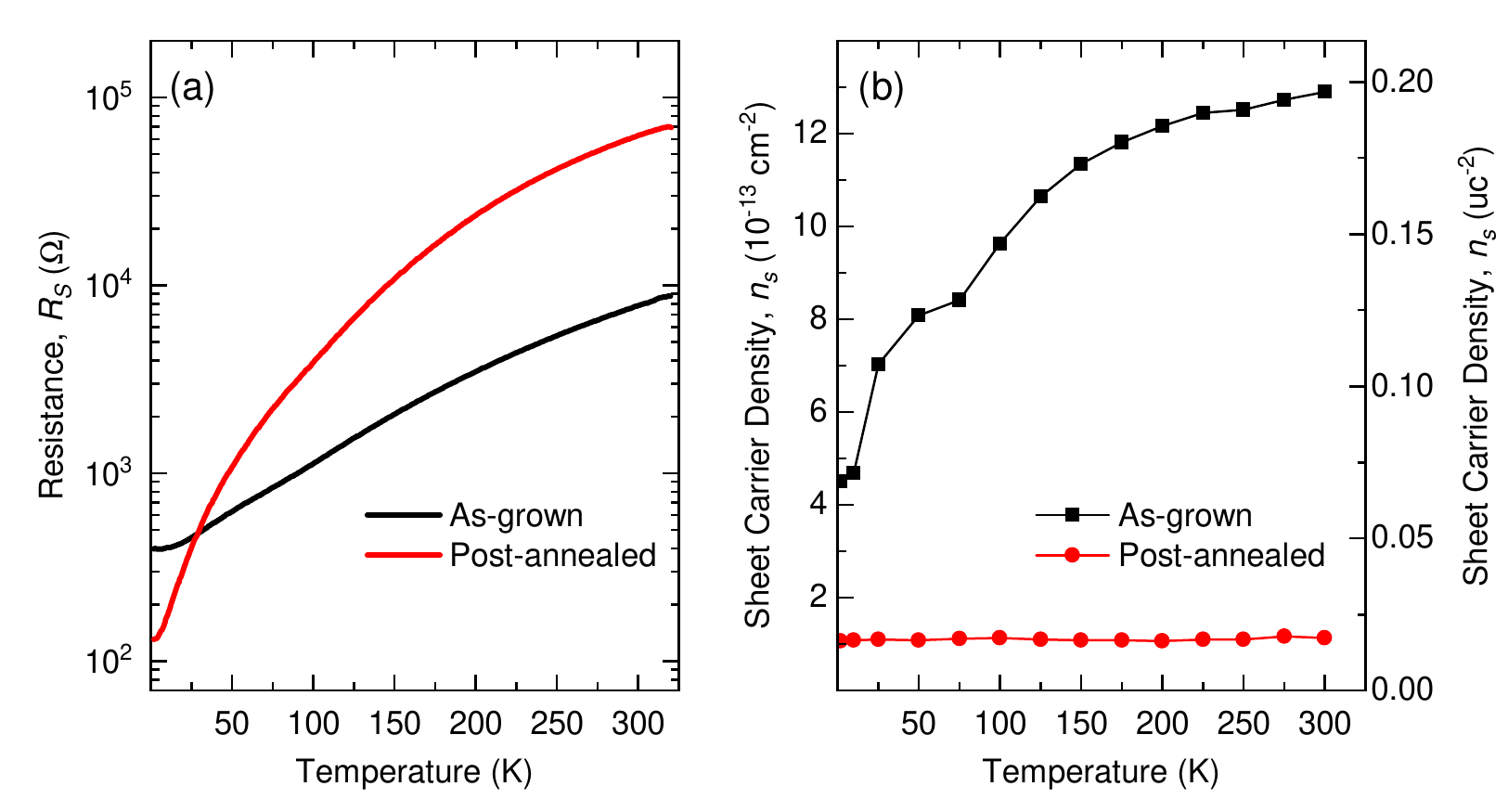}
\caption{\textbf{Transport properties for 10 u.c. films.} \textbf{a}, Temperature dependent sheet resistance measured for 10 u.c. samples as-grown and after the O$_2$ post-annealing procedure. \textbf{b}, Sheet carrier density for as-grown and post-annealed samples.}
\label{Fig:Transport}
\end{center}
\end{figure*}

After a resonant x-ray reflectometry (RXR) dataset was collected for the 10 u.c. sample, it was removed from the x-ray scattering chamber and immediately annealed at 530$^\circ$C in 200 mbar of O$_2$ for 1 hour. The sample was then cooled, and reintroduced into the x-ray scattering chamber for collection of a second RXR dataset.  Transport characteristics for similarly prepared and annealed samples were measured and are shown in Figure \ref{Fig:Transport}. The resistance, shown in Fig. \ref{Fig:Transport}\textcolor{blue}{a}, shows that the as-grown sample has higher conductivity at room temperature, but not at low temperatures.  This is understandable from the carrier density shown in Fig. \ref{Fig:Transport}\textcolor{blue}{b}, where the as-grown sample has considerably more carriers at room temperature, but due to thermal activation loses a considerable fraction of these carriers at low temperature. At 300 K (the temperature of the RXR experiments), the as grown sample shows roughly 0.20 e$^-$/u.c., where-as after annealing this number drops to roughly an order of magnitude to 0.02 e$^-$/u.c. These numbers can be compared to the RXR charge concentrations of 0.97 and 0.50 e$^-$/u.c. in the as-grown and annealed sample, respectively. Evident is that RXR detects more charge in general, understandable since RXR detects both localized and mobile carriers and the total charge decreases by roughly half upon annealing whereas mobile charge decreases by an order of magnitude. Put differently, for the as-grown sample with oxygen vacancies, about 20\% of the charge is mobile, whereas for the annealed system presumably exhibiting pure electronic reconstruction, about 4\% of the charge is mobile.

\section{Reflectivity Measurement Details}

The RXR measurements were performed at the Resonant Elastic and Inelastic X-ray Scattering (REIXS) beamline of the Canadian Light Source \cite{Hawthorn_REIXS_Supp}. An elliptically polarizing undulator (EPU) provided the vertical and horizontal linearly polarized x-rays. An in-vacuum diffractometer accommodated the reflection-geometry scans, and a photodiode was used to detect the reflected beam intensity. The response function of the photodiode was measured using the direct synchrotron beam.  Data was normalized by the incident beam flux and the photodiode response to obtain the quantitative reflectivity spectra. During the experiments, the energy resolution E/$\Delta$E was \mytilde $10^4$, the sample temperature during measurement was 300 K, and the pressure in the measurement chamber was $\lesssim 10^{-9}$ Torr.

Compared to other x-ray scattering measurements (fluorescence yield x-ray absorption and resonant inelastic x-ray scattering, for example), the reflectivity signal is very strong over the $q_z$ ranges studied here. Accordingly, we were able to reduce the beam flux incident upon the sample considerably compared to other measurement techniques, thus minimizing any possibility of radiation damage to our samples. The estimated beam intensity on our samples during measurements is \mytilde 10$^{10}$ photons/second over the area of the \mytilde (0.5 $\times$ 0.2) mm$^2$ beam cross section.

\section{Resonant Scattering Tensor Calculations}

The atomic scattering factors which make up the dielectric response of our films are tabulated for off-resonant energies \cite{Henke_Supp}. The tabulated factors are not accurate at the resonances, however, and therefore we explicitly calculate the Ti resonance response which enters into our reflectivity model.

The $L_{2,3}$ absorption resonances of transition metal oxides have been generally understood for more than 30 years \cite{DeGroot_PRB_1990}. In many cases, the response is dominated by multiplet peaks and can be accurately calculated with quantum many-body multiplet crystal field theory (MCFT).  This model Hamiltonian approach includes the full multiplet effects arising from strong atomic Coulomb interactions, as well as crystal field perturbations arising from the bonding with oxygen. Exact diagonalization and simulations of the spectral functions were carried out with the software \emph{Quanty}.\cite{QuantyWeb_Supp, Haverkort_Wannier_PRB_2010_Supp, Haverkort_DMFT_2014_Supp, Haverkort_DMFTXAS_EPL2014_Supp} The simulated resonances were merged with tabulated off-resonant atomic scattering factors $f_2$,\cite{Henke_Supp} and the scattering factors $f_1$ were then calculated from the Kramers-Kronig relations. For La, the resonant scattering factor was determined from the x-ray absorption spectrum and Kramers-Kronig methods. For all other elements (Sr, O, Al), off-resonant tabulated scattering factors were used. With all atomic scattering factors determined, the reflectivity was simulated and fit to experiment using the dynamical diffraction software QUAD.\cite{QUAD_PRL_Supp}

\begin{figure}[h]
\includegraphics[width=179mm]{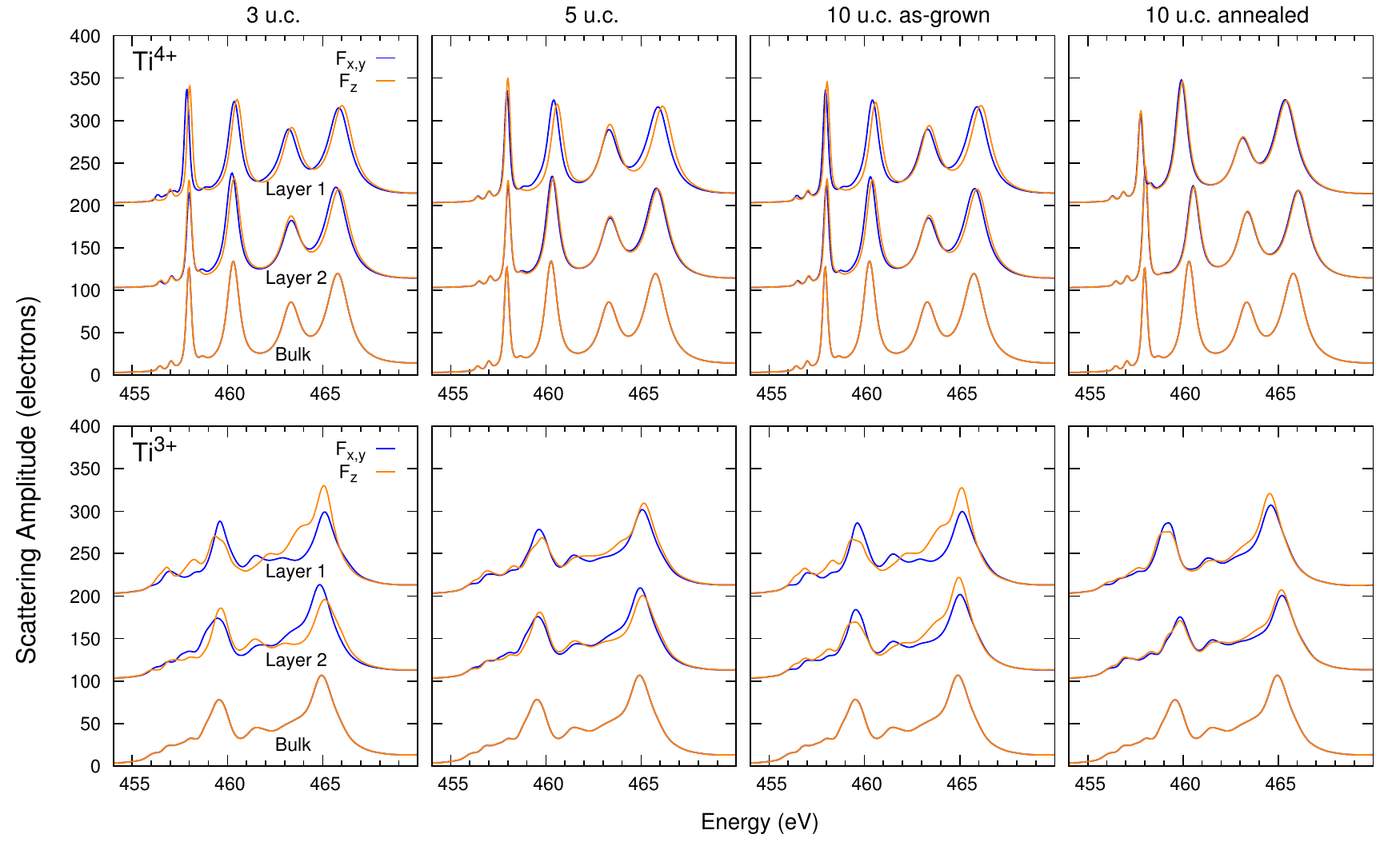}
\caption{\label{Fig:ff} \textbf{Imaginary component of computed Ti $L_{2,3}$ resonant scattering tensors.} (Top row) Ti$^{4+}$ scattering tensors for the first two orbital-reconstructed layers and the bulk of the crystal for the samples having the LaAlO$_3$ thickness indicated.  (Bottom row) similar to the above, but showing the imaginary component of the Ti$^{3}$ scattering tensor.  } 
\end{figure}

As stated, we model the Ti resonant response with MCFT, where the Ti valence and $3d$ orbital energies are the main parameters of the model to be fit to the data. Reliable values for the other parameters of such calculations are well established, and the values used in this work are in Table \ref{Tab:QuantyPars}. In addition to these parameters, the resonant scattering tensors are determined by the orbital energies given in Fig. 3 of the main text. Given the parameters of Table \ref{Tab:QuantyPars}, and the layer-dependent orbital energies given in the main text, the imaginary components of the scattering tensors are as shown in Fig. \ref{Fig:ff} (the real parts are straightforwardly obtained via a Kramers-Kronig transform, and are therefore not shown).

\begin{table}[h]
\begin{minipage}[c]{5.5in}
\begin{ruledtabular}
\begin{tabular}{cccccccccc}
~ & $F^2_{dd}$ & $F^4_{dd}$ & $F^2_{pd}$ & $G^1_{pd}$ & $G^3_{pd}$ & $\zeta_{3d}$ & $\zeta_{2p}$ & $10Dq$ & $\delta t_{2g}(L_2)$ \\ \hline
Ti$^{4+}$ &   -   &   -   & 3.628 & 3.153 & 1.792 & 0.019 & 3.820 & 2.12 & -0.20\\
Ti$^{3+}$ & 5.907 & 3.711 & 3.187 & 3.393 & 1.258 & 0.019 & 3.710 & 1.80 & 0.00  \\
\end{tabular}
\end{ruledtabular}
\end{minipage}
\caption{\label{Tab:QuantyPars} Parameters for quantum many-body model calculations of scattering tensors (all units are eV). $F$ and $G$ parameters are Coulomb and exchange multipole radial integrals, $\zeta$ parameters are spin orbit interaction energies for the indicated shells, and $10Dq$ is the $e_g-t_{2g}$ crystal field splitting. $\delta t_{2g}(L_2)$ is a constant shift added to the $t_{2g}$ $L_2$ peak excitations to account for hybridization effects beyond MCFT. }
\end{table}

\section{Simulations of Reflectivity}

With the atomic scattering factors determined (either from tabulated values or direct calculations), the reflectivity response of the heterostructure can be simulated at various levels of sophistication.  As we are sensitive to, and interested in, the effects at the level of atomic layers, we use the recently developed software QUAD \cite{Macke2016,QUADWeb}, which uses a layered multiple scattering method perfectly suited to thin film heterostructures. A (001) perovskite-specific model is employed, with AO and BO$_2$ alternating planes comprising the ABO$_3$ structure of LaAlO$_3$ and SrTiO$_3$. Intermixing of Al/Ti and Sr/La is included, and the degree of intermixing is fit using an error-function profile. The exact (sub-unit-cell) thickness of the LaAlO$_3$, the surface roughness of the LaAlO$_3$, and the thickness, density, and roughness of an organic (carbon, for simplicity) surface contamination layer due to atmosphere exposure is included. Additionally, while the films were nominally TiO$_2$-terminated, the model allowed for a fractional coverage of SrO or excess TiO$_2$ termination, presented in Table \ref{Tab:RXRPars} as an excess or deficiency in the nominal thickness of the SrO sublattice, $\delta$SrO.

The Levenburg-Marquardt non-linear least squares algorithm is used for fitting the parameters using the experimental RXR data. Various randomized starts as well as a genetic algorithm for initial rough fits are used to ensure that a globally optimized set of parameters is attained. The error values shown are determined from (the square root of) the diagonal of the fit covariance matrix (off diagonal elements in all cases were significantly smaller than diagonal, indicating little correlation of fit parameters). 

\begin{figure}
\includegraphics[width=179mm]{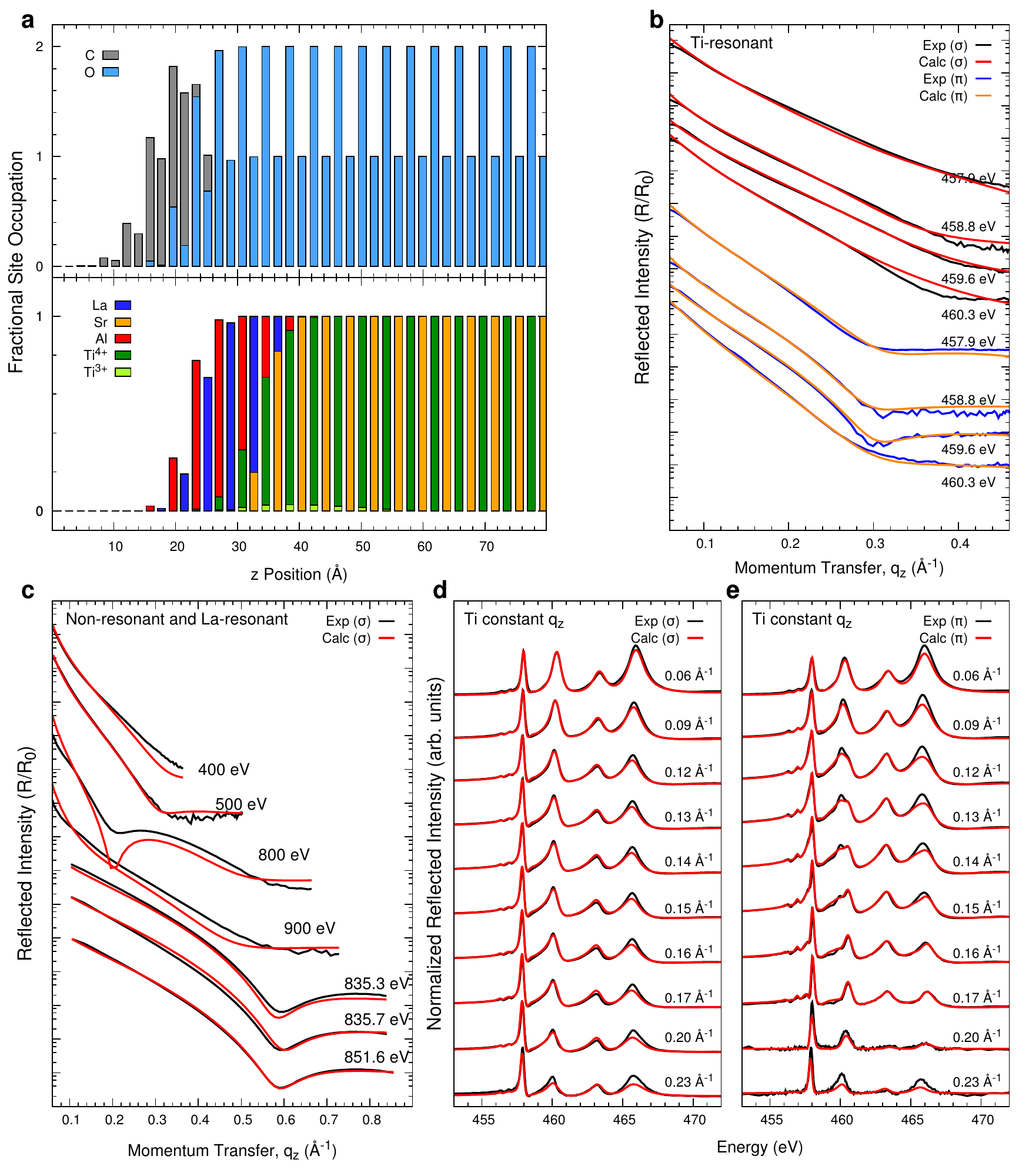}
\caption{\label{Fig:3uc} \textbf{Resonant reflectometry results for 3 u.c. LAO/STO. a}, Atomic layer resolved element and valence concentration depth profiles for anions (top) and cations (bottom). \textbf{b}, Reflectivity angle scans for energies at the Ti $L_{2,3}$ resonance. \textbf{c}, Reflectivity angle scans for non-resonant and La $M_5$-resonant energies. \textbf{d}, Reflectivity energy scans for the indicated momentum transfers using $\sigma$-polarized x-rays. \textbf{e}, Reflectivity energy scans for the indicated momentum transfers using $\pi$-polarized x-rays. Curves are offset vertically for clarity.  } 
\end{figure}

Best fit parameters are shown in Table \ref{Tab:RXRPars}. The extended experimental and simulated data sets are shown on the following pages for each sample, along with detailed plots of the atomic-layer-resolved structures.  Note that in the main manuscript, the depth coordinates of the samples were shifted slightly for presenting aligned interfaces, so the x-axes of Figs. \ref{Fig:3uc}\textcolor{blue}{a}-\ref{Fig:10ucPA}\textcolor{blue}{a} have slightly shifted coordinates in comparison.

\begin{table}[h]
\begin{ruledtabular}
\begin{tabular}[c]{ lcccc }  %\hline \hline
 		~~~~\textbf{Parameter} & \textbf{3 u.c.}  & \textbf{5 u.c.}  & \textbf{10 u.c.}  & \textbf{10 u.c. annealed} \\      \hline
  		\textbf{Thickness} (\AA) &   \\ 
  		~~~~~La & $10.58 \pm 0.05$  & $20.08 \pm 0.05$  & $40.77 \pm 0.06$  & $40.62 \pm 0.10$ \\
  		~~~~~Al & $11.38 \pm 0.41$  & $19.08 \pm 0.53$  & $40.58 \pm 0.25$  & $39.52 \pm 0.37$ \\
  		~~~~~C  & $~2.10 \pm 0.25$  & $~2.98 \pm 0.40$  & $~5.02 \pm 0.51$  & $~2.93 \pm 0.18$ \\
  		~~~~~$\delta$SrO & $~0.07 \pm 0.07$ & $-1.08 \pm 0.10~$ & $-1.38 \pm 0.07~$ & $-0.92 \pm 0.08~$ \\
  		\textbf{Roughness} (\AA) &  \\ 
  		~~~~~Sr/La     & $2.98 \pm 0.01$  & $4.73 \pm 0.06$  & $4.63 \pm 0.10$  & $4.32 \pm 0.07$ \\ 
  		~~~~~Ti/Al     & $5.40 \pm 0.02$  & $5.02 \pm 0.05$  & $4.78 \pm 0.04$  & $5.31 \pm 0.03$  \\
  		~~~~~(La,Al)/C & $3.89 \pm 0.02$  & $4.62 \pm 0.06$  & $3.89 \pm 0.11$  & $4.23 \pm 0.13$  \\
  		~~~~~C/vacuum        & $7.11 \pm 0.15$  & $3.86 \pm 0.30$  & $6.58 \pm 0.36$  & $0.99 \pm 0.90$  \\
  		\textbf{Concentration} (uc$^{-2}$) &  \\ 
  		~~~~~Ti$^{3+}$ & $0.16 \pm 0.03$ & $0.59 \pm 0.05$ & $0.97 \pm 0.08$ & $0.50 \pm 0.03$ \\ 
  		~~~~~C         & $5.49 \pm 0.21$ & $5.26 \pm 0.40$ & $6.80 \pm 0.23$ & $2.68 \pm 0.14$ \\ \hline %\hline%
\end{tabular}
\end{ruledtabular}
\caption{\label{Tab:RXRPars} Best-fit parameters for RXR models of LAO/STO samples.}
\end{table}

\clearpage

\begin{figure}
\includegraphics[width=179mm]{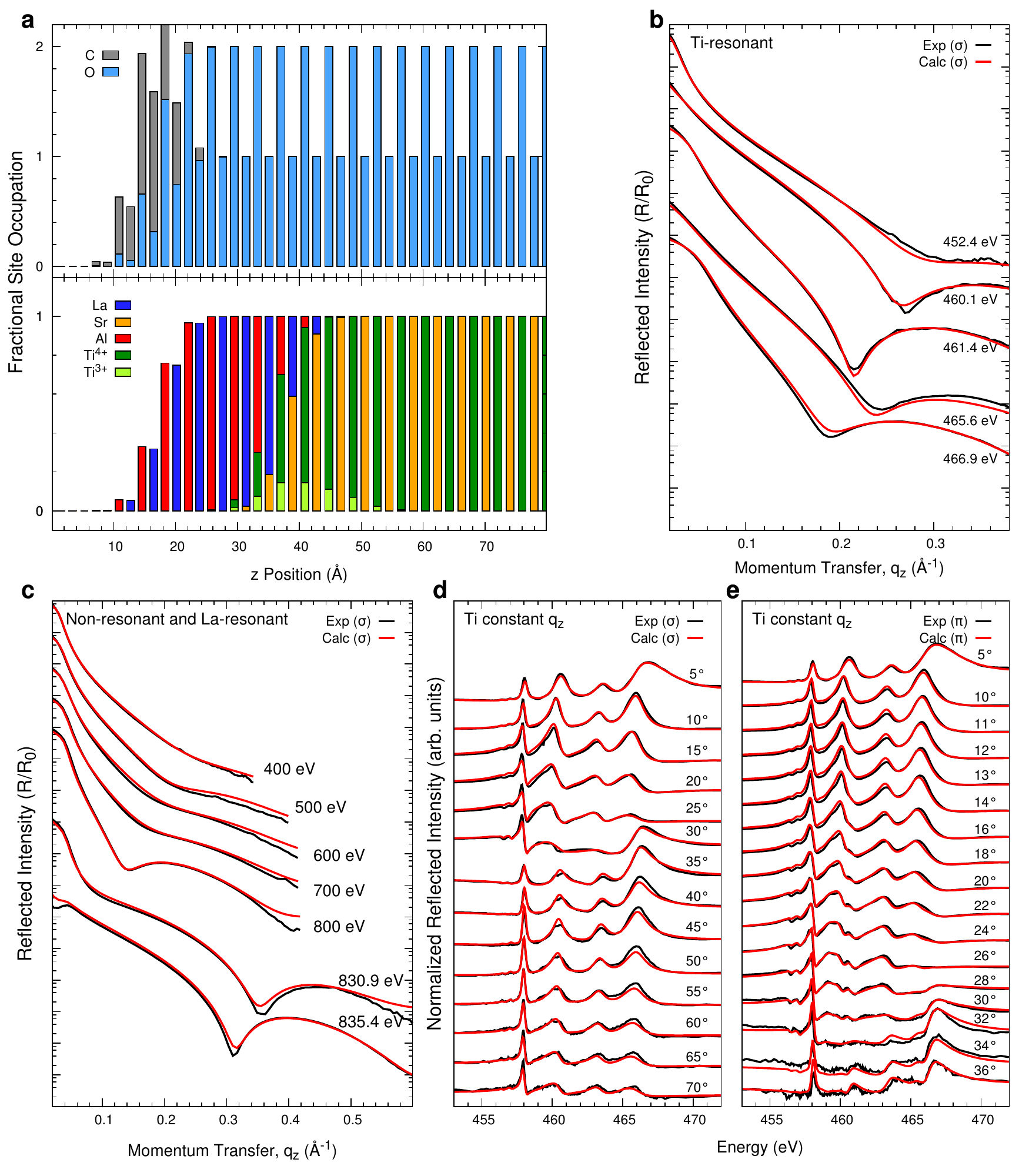}
\caption{\label{Fig:5uc} \textbf{Resonant reflectometry results for 5 u.c. LAO/STO. a}, Atomic layer resolved element and valence concentration depth profiles for anions (top) and cations (bottom). \textbf{b}, Reflectivity angle scans for energies at the Ti $L_{2,3}$ resonance. \textbf{c}, Reflectivity angle scans for non-resonant and La $M_5$-resonant energies. \textbf{d}, Reflectivity energy scans for the indicated reflection angles using $\sigma$-polarized x-rays. \textbf{e}, Reflectivity energy scans for the indicated reflection angles using $\pi$-polarized x-rays. Curves are offset vertically for clarity.  } 
\end{figure}

\begin{figure}
\includegraphics[width=179mm]{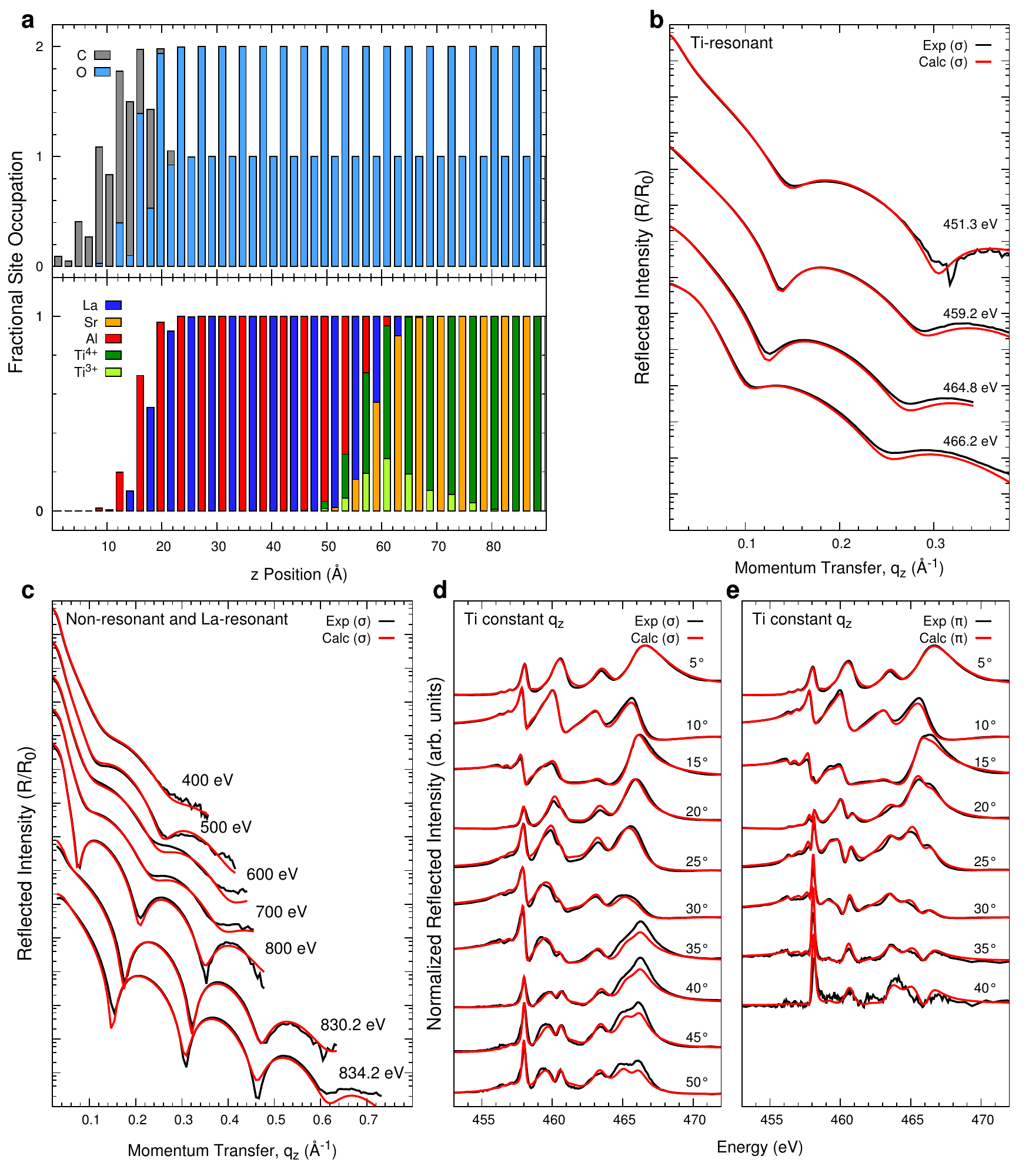}
\caption{\label{Fig:10ucNPA} \textbf{Resonant reflectometry results for 10 u.c. LAO/STO before post-growth annealing. a}, Atomic layer resolved element and valence concentration depth profiles for anions (top) and cations (bottom). \textbf{b}, Reflectivity angle scans for energies at the Ti $L_{2,3}$ resonance. \textbf{c}, Reflectivity angle scans for non-resonant and La $M_5$-resonant energies. \textbf{d}, Reflectivity energy scans for the indicated reflection angles using $\sigma$-polarized x-rays. \textbf{e}, Reflectivity energy scans for the indicated reflection angles using $\pi$-polarized x-rays. Curves are offset vertically for clarity.  }
\end{figure}

\begin{figure}
\includegraphics[width=179mm]{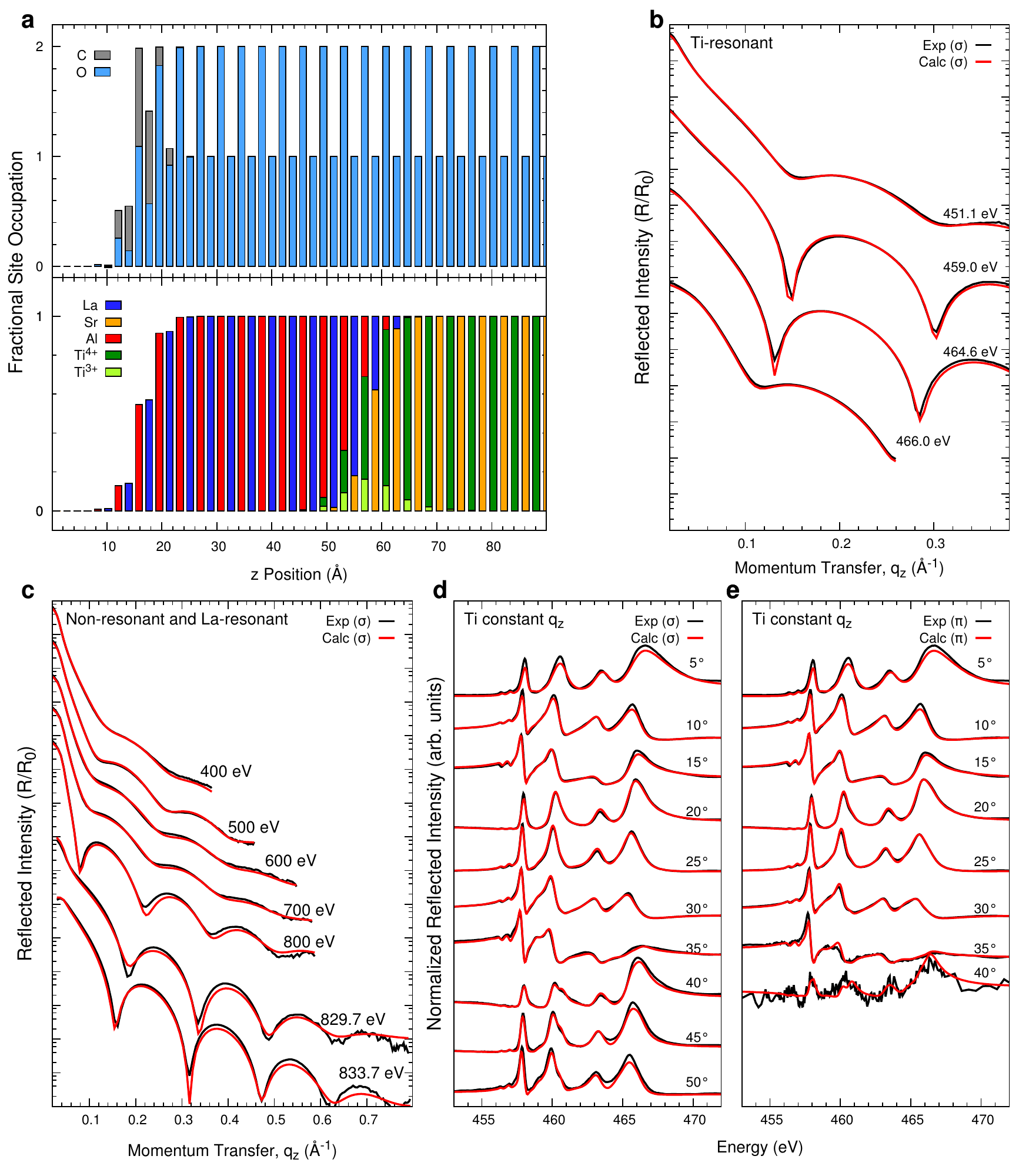}
\caption{\label{Fig:10ucPA} \textbf{Resonant reflectometry results for 10 u.c. LAO/STO after post-growth annealing. a}, Atomic layer resolved element and valence concentration depth profiles for anions (top) and cations (bottom). \textbf{b}, Reflectivity angle scans for energies at the Ti $L_{2,3}$ resonance. \textbf{c}, Reflectivity angle scans for non-resonant and La $M_5$-resonant energies. \textbf{d}, Reflectivity energy scans for the indicated reflection angles using $\sigma$-polarized x-rays. \textbf{e}, Reflectivity energy scans for the indicated reflection angles using $\pi$-polarized x-rays. Curves are offset vertically for clarity.  }
\end{figure}

\end{document}